\documentclass[runningheads]{llncs}

\usepackage[T1]{fontenc}

\usepackage{amsmath}
\usepackage{amssymb}
\usepackage{cite}
\usepackage{booktabs}
\usepackage[inline]{enumitem}
\usepackage{xspace}
\usepackage{xcolor}
\usepackage{pifont}
\usepackage{graphicx}
\usepackage{csquotes}
\usepackage[normalem]{ulem}
\usepackage{bm}

\usepackage{tikz}
\usetikzlibrary{shapes, calc, intersections, shapes.geometric, fit}
\tikzstyle{entity}=[draw,rectangle,rounded corners=.2cm, fill=yellow!20,minimum height=.5cm, minimum width=.75cm,align=center]
\tikzstyle{activity}=[draw,rectangle, fill=blue!20,minimum height=.5cm, minimum width=.75cm,align=center]
\tikzstyle{agent}=[draw,-,agentshape,fill=red!20,minimum height=.5cm, minimum width=.75cm,align=center]
\tikzstyle{use}=[draw,->,>=stealth,edge node={node[fill=white,font=\scriptsize,inner sep=1pt] {\use}}]
\tikzstyle{attr}=[draw,->,>=stealth,edge node={node[fill=white,font=\scriptsize,inner sep=1pt] {\attr}}]
\tikzstyle{deriv}=[draw,->,>=stealth,edge node={node[fill=white,font=\scriptsize,inner sep=1pt] {\deriv}}]
\tikzstyle{gen}=[draw,->,>=stealth,edge node={node[fill=white,font=\scriptsize,inner sep=1pt] {\gen}}]
\tikzstyle{assoc}=[draw,->,>=stealth,edge node={node[fill=white,font=\scriptsize,inner sep=1pt] {\assoc}}]
\tikzstyle{member}=[draw,->,>=stealth,edge node={node[fill=white,font=\scriptsize,inner sep=1pt] {\member}}]

\pgfdeclarelayer{background}
\pgfsetlayers{background,main}

\makeatletter
\pgfdeclareshape{agentshape}{
  \savedanchor\centerpoint{%
    \pgf@x=.5\wd\pgfnodeparttextbox%
    \pgf@y=.5\ht\pgfnodeparttextbox%
    \advance\pgf@y by -.5\dp\pgfnodeparttextbox%
  }
  \inheritsavedanchors[from=rectangle]
  \inheritanchorborder[from=rectangle]
  \savedanchor{\bottomleft}{%
    \pgfpoint{-\pgfshapeinnerxsep}{-\pgfshapeinnerysep}}
  \anchor{south west}{\bottomleft}

  \savedanchor{\bottom}{%
	  \pgfpoint{.5\wd\pgfnodeparttextbox}{-\pgfshapeinnerysep}}
  \anchor{south}{\bottom}

  \savedanchor{\bottomright}{%
    \pgfpoint{\dimexpr\wd\pgfnodeparttextbox+\pgfshapeinnerxsep}{-\pgfshapeinnerysep}}
  \anchor{south east}{\bottomright}

  \savedanchor{\right}{%
	  \pgfpoint{\dimexpr\wd\pgfnodeparttextbox+\pgfshapeinnerxsep}{\dimexpr.5\ht\pgfnodeparttextbox+\pgfshapeinnerysep}}
  \anchor{east}{\right}

  \savedanchor{\topleft}{%
    \pgfpoint{-\pgfshapeinnerxsep}{\dimexpr\ht\pgfnodeparttextbox+\pgfshapeinnerysep}}
  \anchor{north west}{\topleft}

  \savedanchor{\left}{%
	  \pgfpoint{-\pgfshapeinnerxsep}{\dimexpr.5\ht\pgfnodeparttextbox+\pgfshapeinnerysep}}
  \anchor{west}{\left}

  \savedanchor{\topright}{%
    \pgfpoint{\dimexpr\wd\pgfnodeparttextbox+\pgfshapeinnerxsep}{\dimexpr\ht\pgfnodeparttextbox+\pgfshapeinnerysep}}
  \anchor{north east}{\topright}

  \savedanchor{\top}{%
    \pgfpoint{.5\wd\pgfnodeparttextbox}{\dimexpr1.5\ht\pgfnodeparttextbox+\pgfshapeinnerysep}}
  \anchor{north}{\top}

  \backgroundpath{
    \pgfpathmoveto{\bottomleft}
    \pgfpathlineto{\bottomright}
    \pgfpathlineto{\topright}
    \pgfpathlineto{\top}
    \pgfpathlineto{\topleft}
    \pgfpathclose
  }
  \anchor{center}{\centerpoint}
}
\makeatother

\newsavebox{\desktopbox}
\savebox\desktopbox{
\begin{tikzpicture}[outer sep=0,inner sep=0]
  \draw[thick,rounded corners=.1cm,line width=.05cm,fill=white]
    (0,0) -- (.5,0) -- (.5,.7) -- (-.5,.7) -- (-.5,0) -- cycle;
    
  \draw[thick,rounded corners] (0,0) -- (0,-.2);
  \draw[thick,rounded corners] (-.3,-.18) -- (.3,-.18);
\end{tikzpicture}
}

\usepackage{caption,subcaption}
\usepackage{latexsym}

\usepackage{listings}
\usepackage{color}

\usepackage{float}
\newfloat{lstfloat}{htbp}{lop}
\floatname{lstfloat}{Listing}

\usepackage{todonotes}
\colorlet{mscolor}{red!20}
\colorlet{awcolor}{green!20}

\let\SHOWCOMMENTS=t

\ifx\SHOWCOMMENTS t

\newcommand{\commentms}[1]{\todo[color=mscolor!30]{\textbf{MS:} #1}}
\newcommand{\commentinms}[1]{\todo[inline,color=mscolor!30]{\textbf{MS:} #1}}
\newcommand{\todoms}[1]{\todo[color=mscolor]{\textbf{TODO (MS):} #1}}
\newcommand{\todoinms}[1]{\todo[inline,color=mscolor]{\textbf{TODO (MS):} #1}}

\newcommand{\commentaw}[1]{\todo[color=awcolor!30]{\textbf{AW:} #1}}
\newcommand{\commentinaw}[1]{\todo[inline,color=awcolor!30]{\textbf{AW:} #1}}
\newcommand{\todoaw}[1]{\todo[color=awcolor]{\textbf{TODO (AW):} #1}}
\newcommand{\todoinaw}[1]{\todo[inline,color=awcolor]{\textbf{TODO (AW):} #1}}

\newcommand{\discuss}[1]{\todo[inline]{\textbf{Discuss:} #1}}
\newcommand{\todoin}[1]{\todo[inline]{\textbf{TODO:} #1}}
\newcommand{\remove}[1]{\sout{#1}}
\else

\newcommand{\commentms}[1]{}
\newcommand{\commentinms}[1]{}
\newcommand{\todoms}[1]{}
\newcommand{\todoinms}[1]{}

\newcommand{\commentaw}[1]{}
\newcommand{\commentinaw}[1]{}
\newcommand{\todoaw}[1]{}
\newcommand{\todoinaw}[1]{}

\newcommand{\discuss}[1]{}
\newcommand{\todoin}[1]{}
\newcommand{\remove}[1]{}
\fi

\newcommand{\complex}{\mathbb{C}}

\newcommand{\graph}{G}

\newcommand\restr[2]{{
  \left.\kern-\nulldelimiterspace 
  #1 
  \vphantom{\big|} 
  \right|_{#2} 
  }}

\usepackage[outline]{contour}

\newcommand{\ent}{\textsc{Ent}}
\newcommand{\act}{\textsc{Act}}
\newcommand{\agt}{\textsc{Agt}}

\newcommand{\attr}{\textsc{Attr}\xspace}
\newcommand{\deriv}{\textsc{Deriv}\xspace}
\newcommand{\gen}{\textsc{Gen}\xspace}
\newcommand{\use}{\textsc{Use}\xspace}
\newcommand{\assoc}{\textsc{Assoc}\xspace}

\newcommand{\cmark}{\text{\contour{black}{{\color{green!70!black}{\ding{51}}}}}\xspace}%
\newcommand{\ucmark}{\text{\contour{black}{{\color{red!70!black}{\ding{55}}}}}\xspace}%
\newcommand{\qmark}{\textbf{\contour{black}{{\color{yellow!85!black}{?}}}}\xspace}%

\begin{document}

\title{Towards Specificationless Monitoring of Provenance-Emitting Systems}
\author{Martin Stoffers\orcidID{0000-0003-2987-4345} \and Alexander Weinert\orcidID{0000-0001-8143-246X}}
\authorrunning{M. Stoffers and A. Weinert}
\institute{German Aerospace Center (DLR)\\Institute for Software Technology\\Cologne, Germany\\\email{firstname.lastname@dlr.de}}

\maketitle

\begin{abstract}
Monitoring often requires insight into the monitored system as well as concrete specifications of expected behavior.
More and more systems, however, provide information about their inner procedures by emitting provenance information in a W3C-standardized graph format.

In this work, we present an approach to monitor such provenance data for anomalous behavior by performing spectral graph analysis on slices of the constructed provenance graph and by comparing the characteristics of each slice with those of a sliding window over recently seen slices.
We argue that this approach not only simplifies the monitoring of heterogeneous distributed systems, but also enables applying a host of well-studied techniques to monitor such systems.
\end{abstract}

\section{Introduction}
\label{sec:introduction}

In current research on monitoring complex systems, the system is often abstracted to a set of streams of values~\cite{DauerFinkbeinerSchirmer2021}.
Even when monitoring distributed systems, the problem of transporting the distributed data streams to the monitor is assumed to be solved by the monitored system~\cite{MomtazBasnetAbbasEtAl2021}.
In modern real-world systems, in contrast, it is a non-trivial effort to engineer a central component that efficiently consolidates data for monitoring in a heavily distributed system.

Moreover, these streams may be annotated with metadata, e.g., information about their creation times~\cite{NenziBortolussiCianciaEtAl2015}, whether an agent is a natural person or a software agent, or whether some input data was required or optional for the execution of a process.
However, these metadata typically do not include information on their relation.
Consider, e.g., a scenario in which a system provides a value~$x$ as an output to the user and also uses~$x$ as input for further computations.
This monitored information typically does not indicate whether the two values coincide by design or by accident.
Although this relation might be recovered from the logging stream, doing so is typically incomplete and error-prone.
The metadata and relations of data produced by a system are known as provenance data~\cite{W3CProv}.

Monitoring the provenance data of a system addresses both issues identified above:
Provenance data describes, among other information, the relation between individual data points emitted by the system.
Thus, it is typically consolidated by the system itself and made available for inspection or monitoring.
Moreover, it contains more information on the inner workings of the system than the functional data.
Hence, monitoring both provenance data and the functional data may lead to earlier detection of undesired system states.

In practice, monitoring approaches may take metadata of functional data into account.
These metadata, however, are usually domain-specific.
In contrast, provenance data are non-domain-specific, yet provide information about the structure of the monitored data as well as meta-data.

One major challenge when monitoring provenance data, however, is that users typically lack intuition about the relation between data they expect from the system.
Thus, formulating specifications for monitoring provenance data is harder for users than formulating specifications for classical monitoring.

To alleviate this shortcoming, in this work we instead focus on anomaly detection, thus using previously seen data as specification. 
We believe monitoring of provenance data to be an interesting problem. 
The explicit graphs of provenance data allow for the application of well-known graph analyses to monitoring.

We present an approach for monitoring provenance data for anomalies without requiring explicit specifications.
To this end, we proceed as follows:
After discussing related work in Section~\ref{sec:related-work}, we we formally define provenance data as provenance graphs in Section~\ref{sec:provenance-graphs} before subsequently describing the monitoring of provenance-emitting systems in Section~\ref{sec:monitoring}.
Afterwards we outline how to use spectral graph theory to detect anomalous provenance data in Section~\ref{sec:anomaly-detection}.
Finally, in Section~\ref{sec:conclusion} we summarize our work and give an outlook on future work.

\section{Related Work}
\label{sec:related-work}

\paragraph*{Provenance}
Early works highlight the importance of provenance to enable audits of automated workflow systems~\cite{FosterVocklerWildeEtAl2002,LudaescherAltintasBerkleyEtAl2006}.
Moreau identified the building blocks for standardized provenance recording and proposes the Open Provenance Model (OPM)~\cite{Moreau2010}, which was later superseded by the W3C PROV standard~\cite{W3CProv,Moreau2015}.
Provenance data is either extracted from software systems after~\cite{SchreiberBoerKurnatowski2021} or during their operation~\cite{KuehnertGoeddekeHerschel2021,JohnsonParadiesDembskaEtAl2021,StoffersMeinelHofmannEtAl2022}.
There is active work towards recording provenance information without instrumenting the system or process~\cite{GehaniTariq2012,BatesTianButlerEtAl2015,PasquierHanGoldsteinEtAl2017,AlterGagerHeusEtAl2022}.

\paragraph*{System Monitoring}
Runtime verification (or system monitoring) is an established building block for ensuring system correctness~\cite{LeuckerSchallhart2009,BartocciFalconeFrancalanzaEtAl2018}.
Existing approaches to monitoring often take a specification of ``good'' or ``bad'' patterns and efficiently detect them in the output data of the system.
This specification is typically given in temporal logics~\cite{Pnueli1977,Koymans1990,BauerLeuckerSchallhart2011,MalerNickovic2012,DawesReger2019,DawesBianculli2021} or in higher-level languages~\cite{DAngeloSankaranarayananSanchezEtAl2005,FaymonvilleFinkbeinerSchledjewskiEtAl2019,BaumeisterFinkbeinerSchirmerEtAl2020}.

\paragraph*{Anomaly Detection in Provenance Graphs}
Since provenance data give structured information about the relation between data points emitted by the system, there has been work towards identifying anomalies in provenance data.
However, such work explicitly focuses on the detection of attacks on the system~\cite{BerradaCheney2019}, uses bespoke projections from graphs to vector spaces~\cite{HanPasquierEtAl2017,HanPasquierEtAl2018} or only analyzes individual characteristics of the computation process~\cite{PouchardHuckMatyasfalviEtAl2018}.
Other approaches aim to compare provenance graphs by ``summarizing'' via clustering~\cite{MackoMargoEtAl2013,AlawiniChenEtAl2018}.

\section{Provenance Graphs}
\label{sec:provenance-graphs}

The W3C defines provenance information as \enquote{information about entities, activities, and people involved in producing a piece of data or thing, which can be used to form assessments about its quality, reliability or trustworthiness.}~\cite{W3CProv}

The PROV standard prescribes a non-domain-specific graph-based ontology of such information.
Each vertex in such a graph denotes either an \emph{entity}, i.e., some piece of data, an \emph{activity}, i.e., a process or action, or an \emph{agent}, i.e., a person, machine or software responsible for a process.
The edges between these vertices denote the relationships between entities, activities, and agents.

\begin{figure}
  \centering
  \begin{tikzpicture}[thick,xscale=3.5,yscale=1.6]
  \node[agent] (agent) at (0,0) {Agent};
  \node[entity] (entity) at (1,.35) {Entity};
  \node[activity] (activity) at (1,-.35) {Activity};

  \path (entity) edge[->,>=stealth,edge node={node[fill=white,font=\scriptsize,inner sep=1pt] {\attr}}] (agent);
  \path (entity) edge[->,>=stealth,out=20,in=-20,looseness=4.5,edge node={node[fill=white,font=\scriptsize,inner sep=1pt] {\deriv}}] (entity);
  \path (entity) edge[->,>=stealth,bend left,edge node={node[fill=white,font=\scriptsize,inner sep=1pt] {\gen}}] (activity);

  \path (activity) edge[->,>=stealth,edge node={node[fill=white,font=\scriptsize,inner sep=1pt] {\use}}, bend left] (entity);
  \path (activity) edge[->,>=stealth,edge node={node[fill=white,font=\scriptsize,inner sep=1pt] {\assoc}}] (agent);
\end{tikzpicture}
  \caption{The core provenance meta-model. Reproduced after and adapted from the Prov Primer~\cite{W3CProvPrimer}.}
  \label{fig:prov-model}
\end{figure}
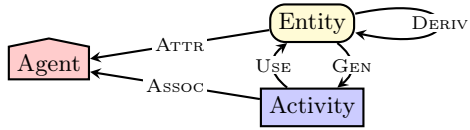

We illustrate the possible relationships in Figure~\ref{fig:prov-model}.
We draw entities as yellow rounded boxes, activities as blue rectangular boxes, and agents as red pentagons.
In Figure~\ref{fig:prov-model}, we write \attr, \deriv, \use, \gen, and \assoc to abbreviate ``was attributed to'', ``was derived from'', ``used'', ``was generated by'', and ``was associated with'', respectively.
The full standard admits additional vertices and edges.
We restrict ourselves to the edges shown in Figure~\ref{fig:prov-model} for conciseness.

Intuitively, the provenance of a software system is a record of
\begin{enumerate*}[label=\alph*)]
  \item the data that was generated or used,
  \item the process that generated and used these data, and
  \item the responsible entities (both human and software) for these processes.
\end{enumerate*}

Formally, a provenance graph~$\graph = (V_\agt, V_\ent, V_\act, E)$ consists of finite sets of agent vertices $V_\agt$, entity vertices $V_\ent$, and activity vertices $V_\act$, all of which are pairwise disjoint, and a set of edges
$E \subseteq (V_\ent \times (V_\agt \cup V_\ent \cup V_\act)) \cup (V_\act \times (V_\agt \cup V_\ent))$.
We call $V_\agt \cup V_\ent \cup V_\act$ the \emph{vertex set} of~$\graph$.

In practice, a provenance-emitting system does not provide its complete provenance at the end of its computation.
Instead, whenever an activity has terminated, the system provides a ``partial'' provenance graph that contains the respective activity as well as the entities that this activity used and generated.
We also call these provenance graphs emitted during execution of the system \emph{provenance updates} to differentiate them from the complete graph that the system constructs during its execution.
To obtain a less local view of the provenance of the complete system execution, we construct the union over provenance via the component-wise union of the constituent elements of a provenance graph.

In this work, a \emph{provenance-emitting system} is a software or hardware system that constructs a provenance graph that contains vertices representing the data points it generated, the processes that used and generated these data points, and makes this provenance graph accessible to external systems.
In the following section, we provide greater detail and a more formal description of such systems.
We moreover describe a process for monitoring provenance-emitting systems.

Having given a brief introduction to the W3C Provenance standard, we now illustrate how provenance data is collected from distributed provenance-emitting systems and made available for monitoring in the following section.

\section{Monitoring Provenance-Emitting Systems}
\label{sec:monitoring}

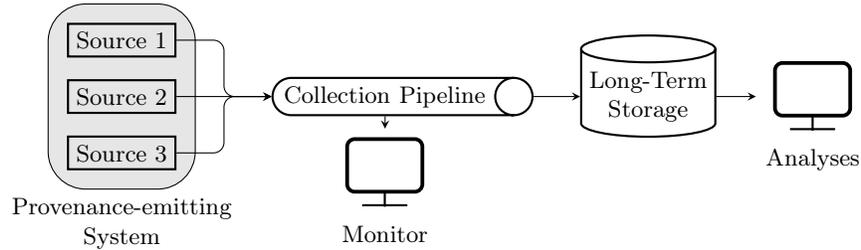
\begin{figure}
  \centering
  \begin{tikzpicture}[thick,xscale=1.75,yscale=0.75]

  \node[draw,align=center] (source-1) at (0, 1) {Source 1};
  \node[draw,align=center] (source-2) at (0, 0) {Source 2};
  \node[draw,align=center] (source-3) at (0, -1) {Source 3};

  \begin{pgfonlayer}{background}
    \node[fit=(source-1) (source-2) (source-3), rounded corners=.5cm, inner sep=.25cm,draw=black, fill=black!10,label={[align=center]below:Provenance-emitting\\System}] {};
  \end{pgfonlayer}

  \node[draw,cylinder,thick] (pipeline) at (2,0) {Collection Pipeline};

  \node[draw,thick,cylinder,shape border rotate=90,aspect=.2,align=center] (storage) at (4, 0) {Long-Term\\Storage};

  \path[draw,->,>=stealth,rounded corners] (source-1) -| ($(source-1.east) ! .5 ! (pipeline.west)$) |- (pipeline);
  \path[draw,->,>=stealth,rounded corners] (source-2) -- (pipeline);
  \path[draw,->,>=stealth,rounded corners] (source-3) -| ($(source-3.east) ! .5 ! (pipeline.west)$) |- (pipeline);

  \node[label=below:Monitor] (monitor) at (2, -1.35) {\usebox{\desktopbox}};

  \path[draw,->,>=stealth] (pipeline) -- (monitor);

  \path[draw,->,>=stealth] (pipeline) -- (storage);

  \node[label={below:Analyses}] (analyses) at (5.25, 0) {\usebox{\desktopbox}};

  \path[draw,->,>=stealth] (storage) -- (analyses);
\end{tikzpicture}
    \caption{A lightweight architecture for capturing provenance information from a complex distributed system. (Taken from~\cite{StoffersMeinelHofmannEtAl2022} and simplified.)}
  \label{fig:provenance-architecture}
\end{figure}

Intuitively, a provenance-emitting system emits information about its execution by making its provenance graph available for monitoring and inspection.
To do so, it emits provenance updates at certain points in time, i.e., sub-graphs of the complete provenance graph of its execution.
A monitor can observe these updates and monitor them for anomalies.
In Figure~\ref{fig:provenance-architecture} we illustrate a lightweight architecture for recording provenance, initially outlined in~\cite{StoffersMeinelHofmannEtAl2022}.

Formally, we say that an \emph{execution} of a provenance-emitting system is an infinite sequence $\graph_1 \graph_2 \graph_3 \cdots$.
The graphs are not necessarily temporally ordered.
As an example, $\graph_3$ may only contain activities that started at time~$t$, while~$\graph_4$ only contains activities that ended at time~$t' < t$.

In this section we present a method to monitor an execution of provenance-emitting systems for anomalies.
The complete provenance graph constructed by the system is unbounded and may be infinite for non-terminating systems.
It is the task of the long-term Storage (cf. Figure~\ref{fig:provenance-architecture}) to provide sufficient storage capacity for this complete graph as required for post-hoc analyses.
The monitor, in contrast, should not require unbounded memory, but instead work with restricted resources to function as lightweight as possible.

\begin{quote}
  Let~$\varphi = \graph_1\graph_2\graph_3 \cdots$ be the execution of a provenance-emitting system.
  In each time step the monitor obtains the earliest~$\graph$ from from~$\varphi$ that it has not yet obtained.
  The monitor shall produce a sequence $b_1 b_2 b_3 \cdots$ where~$b_i$ is one of $\cmark, \ucmark, \qmark$.
  The value~$\cmark$~($\ucmark$) denotes that the monitor has not (has) detected an anomaly in the last time step, respectively, while~$\qmark$ denotes that the monitor does not have sufficient information to make a decision.
  Moreover, the monitor shall not require unbounded memory.
\end{quote}

In this problem formulation, we explicitly omit a definition of ``anomalies''.
Recall that the PROV meta model only imposes very limited structure on provenance graphs.
Thus, whether a provenance graph describes expected or unexpected behavior is strongly application-dependent.
Analogously, it is not expedient to formally define anomalies independent of the monitored application.

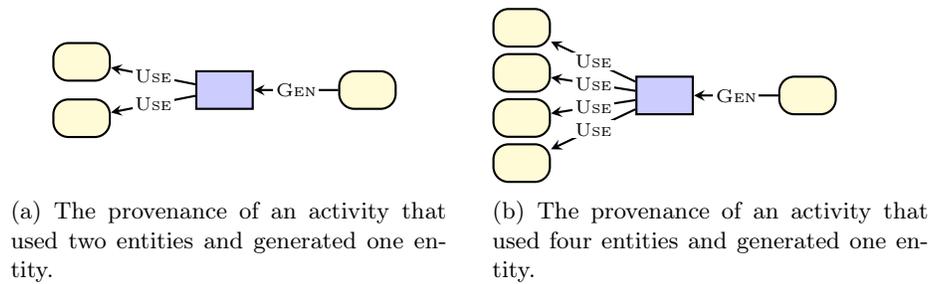
\begin{figure}
  \centering
  \begin{subfigure}[b]{.475\textwidth}
    \centering
    \begin{tikzpicture}[thick,xscale=1.9,yscale=.75]
  \node at (0,-1.5) {};
  \node at (0,1.5) {};

  \node[entity] (data-in-1) at (0,.5) {};
  \node[entity] (data-in-2) at (0,-.5) {};

  \node[activity] (activity) at (1, 0) {};

  \node[entity] (data-out) at (2, 0) {};

  \path
    (activity) edge[use] (data-in-1)
    (activity) edge[use] (data-in-2)
    (data-out) edge[gen] (activity);

\end{tikzpicture}
    \caption{The provenance of an activity that used two entities and generated one entity.}
    \label{fig:anomaly-example:1}
  \end{subfigure}
  \hfill
  \begin{subfigure}[b]{.475\textwidth}
    \begin{tikzpicture}[thick,xscale=1.9,yscale=.6]
  \node[entity] (data-in-1) at (0,1.5) {};
  \node[entity] (data-in-2) at (0,.5) {};
  \node[entity] (data-in-3) at (0,-.5) {};
  \node[entity] (data-in-4) at (0,-1.5) {};

  \node[activity] (activity) at (1, 0) {};

  \node[entity] (data-out) at (2, 0) {};

  \path
    (activity) edge[use] (data-in-1)
    (activity) edge[use] (data-in-2)
    (activity) edge[use] (data-in-3)
    (activity) edge[use] (data-in-4)
    (data-out) edge[gen] (activity);

\end{tikzpicture}
    \caption{The provenance of an activity that used four entities and generated one entity.}
    \label{fig:anomaly-example:2}
  \end{subfigure}
  \caption{Examples of provenance graphs}
  \label{fig:anomaly-example}
\end{figure}

Consider, e.g., a provenance-emitting system in which all activities so far take two entities as inputs and produce one entity as output.
Assume there arrives a provenance update in which an activity takes four inputs and produces one output.
We illustrate the former and latter case in Figure~\ref{fig:anomaly-example:1} and in Figure~\ref{fig:anomaly-example:2}, respectively.
Whether or not the latter update is anomalous depends on the purpose of the system.
If the system processes temperature readings since the last step, then the occurrence of fewer temperature readings than previously indicates, e.g., faulty sensors.
In contrast, if a system processes observations made by an optical telescope~\cite{FiedlerHerzogProhaskaEtAl2017,StoffersMeinelWeigelEtAL2019}, then the system may process more information due to improved weather conditions.
This would not be considered anomalous.

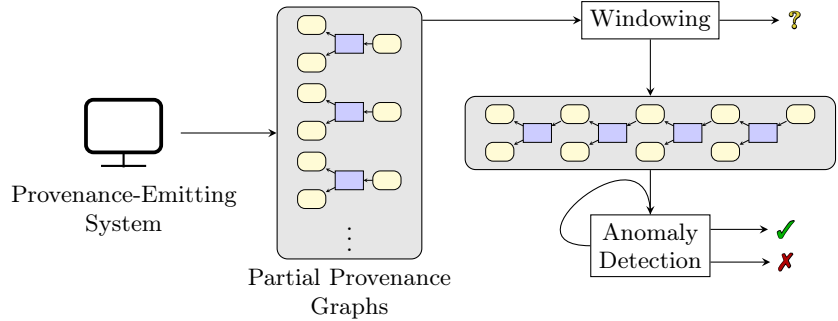
\begin{figure}
  \centering
  \newsavebox{\partialgraph}
\savebox\partialgraph{
\scalebox{.5}{
\begin{tikzpicture}[outer sep=0,inner sep=0]
    \node[entity] (data-in-1) at (0,.5) {};
    \node[entity] (data-in-2) at (0,-.5) {};

    \node[activity] (activity) at (1, 0) {};

    \node[entity] (data-out) at (2, 0) {};

    \path
      (activity) edge[->,>=stealth] (data-in-1)
      (activity) edge[->,>=stealth] (data-in-2)
      (data-out) edge[->,>=stealth] (activity);
\end{tikzpicture}
} }

\newsavebox{\completegraph}
\savebox\completegraph{
\scalebox{.5}{
\begin{tikzpicture}[outer sep=0,inner sep=0]
    \node[entity] (data-1) at (0,.5) {};
    \node[entity] (data-2) at (0,-.5) {};

    \node[activity] (activity) at (1, 0) {};

    \path
      (activity) edge[->,>=stealth] (data-1)
      (activity) edge[->,>=stealth] (data-2);

    \node[entity] (data-1) at (2, .5) {};

    \path
      (data-1) edge[->,>=stealth] (activity);

    \node[entity] (data-2) at (2, -.5) {};

    \node[activity] (activity) at (3, 0) {};

    \path
      (activity) edge[->,>=stealth] (data-1)
      (activity) edge[->,>=stealth] (data-2);

    \node[entity] (data-1) at (4, .5) {};

    \path
      (data-1) edge[->,>=stealth] (activity);

    \node[entity] (data-2) at (4, -.5) {};

    \node[activity] (activity) at (5, 0) {};

    \path
      (activity) edge[->,>=stealth] (data-1)
      (activity) edge[->,>=stealth] (data-2);

    \node[entity] (data-1) at (6, .5) {};

    \path
      (data-1) edge[->,>=stealth] (activity);

    \node[entity] (data-2) at (6, -.5) {};

    \node[activity] (activity) at (7, 0) {};

    \path
      (activity) edge[->,>=stealth] (data-1)
      (activity) edge[->,>=stealth] (data-2);

    \node[entity] (data-1) at (8, .5) {};

    \path
      (data-1) edge[->,>=stealth] (activity);
\end{tikzpicture}
} }

\begin{tikzpicture}
  \node[label={[align=center]below:Provenance-Emitting\\System}] (system) at (0,0) {\usebox{\desktopbox}};
  \node[draw=black,fill=black!10,label={[align=center]below:{Partial Provenance\\Graphs}},align=center,rounded corners] (partial-graphs) at (3,0)
    {\usebox{\partialgraph}\\[.3\baselineskip] \usebox{\partialgraph}\\[.3\baselineskip] \usebox{\partialgraph}\\[0] \vdots};
  \node[draw] (windowing) at (7,1.5) {Windowing};
  \node[draw=black,fill=black!10,rounded corners] (complete-graph) at (7,0) {\usebox{\completegraph}};
  \node[draw,align=center] (anomaly-detection) at (7,-1.5) {Anomaly\\Detection};

  \node (cmark) at ($(anomaly-detection.north east) + (1,-.2)$) {\cmark};
  \node (ucmark) at ($(anomaly-detection.south east) + (1,.2)$) {\ucmark};
  \node (qmark) at ($(windowing.east) + (1,0)$) {\qmark};

  \path
    (system) edge[->,>=stealth] (partial-graphs)
    (partial-graphs.east |- windowing.west) edge[->,>=stealth] (windowing)
    (windowing) edge[->,>=stealth] (complete-graph)
    (complete-graph) edge[->,>=stealth] (anomaly-detection)
    (anomaly-detection) edge[out=180,in=90,looseness=3.5] (anomaly-detection)
    (anomaly-detection.east |- cmark) edge[->,>=stealth] (cmark)
    (anomaly-detection.east |- ucmark) edge[->,>=stealth] (ucmark)
    (windowing) edge[->,>=stealth] (qmark);

\end{tikzpicture}
  \caption{An overview over our monitoring architecture.}
  \label{fig:monitor-architecture}
\end{figure}

This example illustrates that there can be no ``turnkey'' solution for monitoring provenance-emitting systems:
Either the user provides an explicit specification of expected or unexpected provenance patterns, or the monitor requires knowledge about the purpose of the system to infer anomalous graphs itself.
Thus, we aim to construct a parametrized monitor that measures the ``anomaly'' of a provenance update against those updates witnessed previously.
Due to space constraints, this monitor cannot store all previously witnessed provenance graphs explicitly.
Instead, it retains only a window of previously witnessed partial provenance graphs.
We illustrate our monitoring architecture in Figure~\ref{fig:monitor-architecture}.

The main purpose of the windowing is to construct a sequence of classical graphs that it then passes to anomaly detection.
By retiring vertices representing entities that have not been used by the system, windowing ensures that the graphs passed to anomaly detection do not exceed a given size.
We call this size the \emph{window size}.
This allows the anomaly detection to focus on comparing an incoming graph to the one previously obtained and to raise an alarm to the user if these two graphs differ significantly.
To give the anomaly detection enough data to reliably identify structural anomalies, the windowing step reports \qmark until it has collected sufficient provenance updates to fill a predetermined window size.

When converting provenance graphs into classical directed graphs, structural information about the ``kinds'' of vertices is lost, as a classical graph does not differentiate between, e.g., vertices representing processes and vertices representing entities.
Moreover, similar information about the kinds of edges is lost as well.
This information can be reconstructed in the fragment of provenance graphs used in this work.
Such a reconstruction is, however, not necessarily possible when using the full expressive power of W3C Provenance.

To retain this information, it may be encoded as vertex- or edge-weights.
How weights should be assigned is again strongly application-specific and strongly influences the subsequent anomaly detection.
One could, e.g., assign a high weight to all edges adjacent to activity-vertices.
This would lead to the anomaly detection being highly sensitive to anomalous patterns in the vicinity of activity-vertices and less sensitive to the vicinity of other vertices.
In the next section we describe possible approaches to detecting anomalies via spectral graph theory.

\section{Anomaly Detection with Spectral Graph Theory}
\label{sec:anomaly-detection}

To identify structural anomalies of graphs, we need to quantify their topological properties, e.g., patterns of connectivity.
We propose using spectral graph theory~\cite{Spielman2011} to this end.
This theory relies on studying the Eigenvalues and Eigenvectors of matrices associated with graphs, e.g., the adjacency matrix, the degree matrix, or the Laplace matrix.
Intuitively, these values capture the topological properties of the investigated graphs~\cite{Spielman2011}.
Spectral graph theory has been successfully applied in some fields~\cite{Spielman2007, ShiMalik1997}.
In particular,  Gera, Alonso, Crawford, et~al.\, have used spectral graph theory to determine whether incoming observations of a graph significantly deviate from previous observations~\cite{GeraAlonsoCrawfordEtAl2018}.

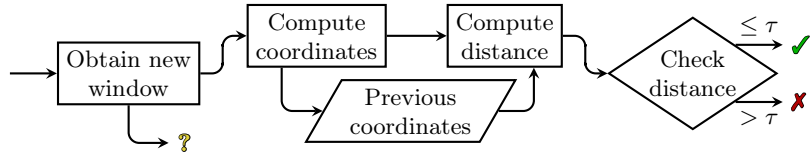
\begin{figure}
  \centering
  \begin{tikzpicture}[thick,xscale=1.25,>=stealth]
  \node[draw,align=center] (obtain) at (0,0) {Obtain new\\window};
  \node[draw,align=center] (compute-coord) at (2,.5) {Compute\\coordinates};
  \node[draw,align=center,trapezium,trapezium left angle=60,trapezium right angle=120] (prev-coord) at (3,-.5) {Previous\\coordinates};
  \node[draw,align=center] (compute-dist) at (4,.5) {Compute \\ distance};
  \node[draw,diamond,aspect=1.5,align=center,inner sep=1pt] (distance-check) at (6,0) {Check \\ distance};

  \node[anchor=east] (qmark) at ($(obtain.south east) + (0,-.5cm)$) {\qmark};
  \node[anchor=west] (cmark) at ($(distance-check.north east) + (.5cm,0)$) {\cmark};
  \node[anchor=west] (ucmark) at ($(distance-check.south east) + (.5cm,0)$) {\ucmark};

  \path ($(obtain.west) - (.5cm,0)$) edge[->] (obtain);
  \path[draw,->,rounded corners] (obtain) -| ($(obtain) ! .5 ! (compute-coord)$) |- (compute-coord);
  \path (compute-coord) edge[->] (compute-dist);
  \path[draw,->,rounded corners] ($(compute-coord.south west) ! .5 ! (compute-coord.south)$) |- (prev-coord);
  \path[draw,->,rounded corners] (prev-coord) -| ($(compute-dist.south) ! .5 ! (compute-dist.south east)$);
  \path[draw,->,rounded corners] (compute-dist) -| ($(compute-dist.east) ! .5 ! (distance-check.west)$) |- (distance-check);

  \path[draw,->,rounded corners] (obtain) |- (qmark);

  \path (distance-check.north east) edge[->] node[anchor=south] {$\leq \tau$} (cmark);
  \path (distance-check.south east) edge[->] node[anchor=north] {$> \tau$} (ucmark);

\end{tikzpicture}
  \caption{Our framework for anomaly detection.}
  \label{fig:anomaly-detection}
\end{figure}

We illustrate our general approach in Figure~\ref{fig:anomaly-detection}.
Let~$\graph$ be an incoming graph obtained by anomaly detection and let~$n$ be the window size determined when constructing or configuring the windowing.
By computing Eigenvalues and Eigenvectors of matrices associated with~$G$ we can obtain vectors~$\nu_1, \dots, \nu_n$, where~$\nu_i \in \complex^m$ for all~$\nu_i$ and some~$m \leq n$.
These vectors may, e.g., comprise the Eigenvalues of the used matrix or its Eigenvectors.
In the former case, we have~$m=1$, in the latter~$m=n$.
Intuitively, if the investigated matrix is well-chosen, this set of vectors quantifies the structure of the graph.
We call this set of vectors the \emph{coordinates} of the graph.
Having obtained the coordinates of both the current and the previous provenance graph, we can then compute the distance between these two coordinates and use this distance as a measure of the structural differences between the two graphs.
We discuss possibilities and challenges for both steps, obtaining coordinates and computing their distance in the following sections.
Moreover, we report on the results of a preliminary evaluation in Section~\ref{sec:anomaly-detection:evaluation}.

\subsection{Compute Coordinates}
\label{sec:anomaly-detection:coordinates}

To obtain coordinates, we compute Eigenvalues and Eigenvectors of matrices associated with the graph.
Typically, one uses the adjacency matrix or the Laplacian matrix of the graph~\cite{Spielman2011}.
Multiple works have shown that some graph properties can be determined based on the multiplicity, size or position of the Eigenvalues and corresponding Eigenvectors~\cite{vonLuxburg2007,BiyikoguLeydoldStadler2007, Mohar1992, Merris1994, Hong1993}.
In graph drawing, Eigenvectors are selected based on their Eigenvalues and used as source for coordinates to visually reveal structural properties of graphs~\cite{Koren2005}.

Most applications of spectral graph theory, however, assume the graph to be undirected.
In that case, the adjacency matrix and the Laplacian matrix are real symmetric matrices, thus their Eigenvalues are integers.
Provenance graphs, however, are directed.
Thus, to apply standard methods of spectral graph analysis to them, we have to transform them into undirected graphs~\cite{SatuluriParthasarathy2011}, which loses structural information.
Another approach would be to use bespoke spectral graph analysis methods that handle directed graphs~\cite{Chung2005,VanLierde2015}.
These methods are, however, not as well-investigated as those for undirected graphs.

\subsection{Compute Distance}
\label{sec:anomaly-detection:distance}

To identify anomalous updates we need to compare the current and the previous coordinate vectors.
To this end, we aim to compute a normalized distance measure.
Recall that coordinates are sets of vectors.
Thus, a common method is to first calculate the pairwise distances for sets of vectors separately using different metrics, such as the euclidean distances, cosine similarity or correlation.
By taking the average over the resulting vector of distance measurements we can obtain a distance between coordinates.

Directly computing the difference between two sets of vertices is rather sensitive to ``noisy'' coordinates:
Minor differences between individual vectors may lead to large differences.
We can counteract this via clustering the vectors comprising the coordinates prior to distance calculation.
In this case, the complete monitoring pipeline up to the computation of coordinates is tantamount to spectral clustering~\cite{vonLuxburg2007} of the provenance graph.
Via clustering we obtain cluster centroids, the coordinates of which can be used to calculate a distance measurement as outlined before.
Having obtained a distance measure, we can check the distance against a provided threshold.
If the distance exceeds this threshold, anomaly detection alerts the user to anomalous system behavior.

\subsection{Proof of Concept}
\label{sec:anomaly-detection:evaluation}

We implemented our method using the Eigenvalues of the Laplacian matrix as coordinates and the distance between the centroids of the Eigenvalues as the distance metric.
We evaluated this implementation on a synthetic example as well as a realistic one.
In the synthetic example, in each time step the system adds two numbers during normal operation.
We have injected anomalies into the provenance data representing the addition of ten numbers in one time step.
In the realistic example, in each time step a robot executes some actions based on some plan~\cite{Huynh2022}.
There are anomalies where no such plan is present.

Our prototypical implementation was able to successfully differentiate between the nominal and anomalous updates.
This illustrates that our proposed method can indeed determine anomalies at least in these two use cases.

\section{Conclusion and Future Work}
\label{sec:conclusion}

In contrast to streams solely comprising the output data of a system, provenance data allows far greater insight into the inner workings of such a system.
Thus, we believe that monitoring provenance data in addition to output data allows for earlier detection of system failures.
We have outlined an approach to monitor these provenance data that reduces the problem of monitoring provenance data to that of determining anomalies in graphs.
This greatly reduces the parameter space that has to be explored when constructing a real-world monitor to determining useful well-studied approaches for detecting anomalies in graphs~\cite{AkogluTongKoutra2014}.
Moreover, we illustrated that this approach can serve as a framework for detecting anomalies in provenance data via a prototypical proof of concept.

As a next step, we aim to identify real-life use cases in which we can apply and evaluate our approach.
This use case will allow us to compare different definitions of coordinates and distances between coordinates.
Moreover, we will be able to evaluate our proposed approach against other approaches to anomaly detection in graphs~\cite{AkogluTongKoutra2014}.
In addition, we aim to quantify the structure of graphs by additional properties, e.g., their diameter or depth.
Finally, we are looking to compare our approach based on spectral graph analysis against existing machine learning approaches to anomaly detection.

\paragraph*{Acknowledgements}
We gratefully acknowledge suggestions by anonymous reviewers, which have significantly improved this work.

\bibliographystyle{splncs04}
\bibliography{rv22-provenance-graph-anomaly-detection}

\end{document}